# High-Throughput Discovery of Semimetallic Borophenes with Diverse Dirac States Via Transferable Tight-Binding Approach

Yuke Song[1,2], Shifang Li[1,2], Tao Ouyang[1,2], Chao Tang[1,2], Jin Li[1,2*] and Chaoyu He[1,2*]

1. School of Physics and Optoelectronics, Xiangtan University, Hunan 411105, China
2. Hunan Provincial Key Laboratory of Computational Condensed Matter Physics and Quantum Materials Engineering, Hunan 411105, China

*Corresponding author. Email: lijin@xtu.edu.cn (Jin Li), hechaoyu@xtu.edu.cn (Chaoyu He)

**ABSTRACT:** Borophene has attracted extensive interest due to its structural flexibility and emergent topological electronic states. However, semimetallic borophenes hosting robust Dirac states remain rare among the large number of predicted allotropes. Here, we develop a transferable tight-binding framework for planar borophenes and combine it with a graph- and group-theory-based random generation strategy to perform high-throughput screening of 522 borophene candidates. Eight previously unreported semimetallic borophenes are identified, hosting diverse topological band crossings, including type-I and type-III Dirac cones, Dirac nodal lines, and quadratic nodal points. Notably, quadratic nodal-point semimetals are predicted in borophene for the first time. Symmetry analysis reveals crystalline-symmetry-protected Dirac states, while first-principles calculations confirm their dynamical and thermal stability. These findings establish borophene as a versatile platform for engineering emergent Dirac physics in two dimensions.

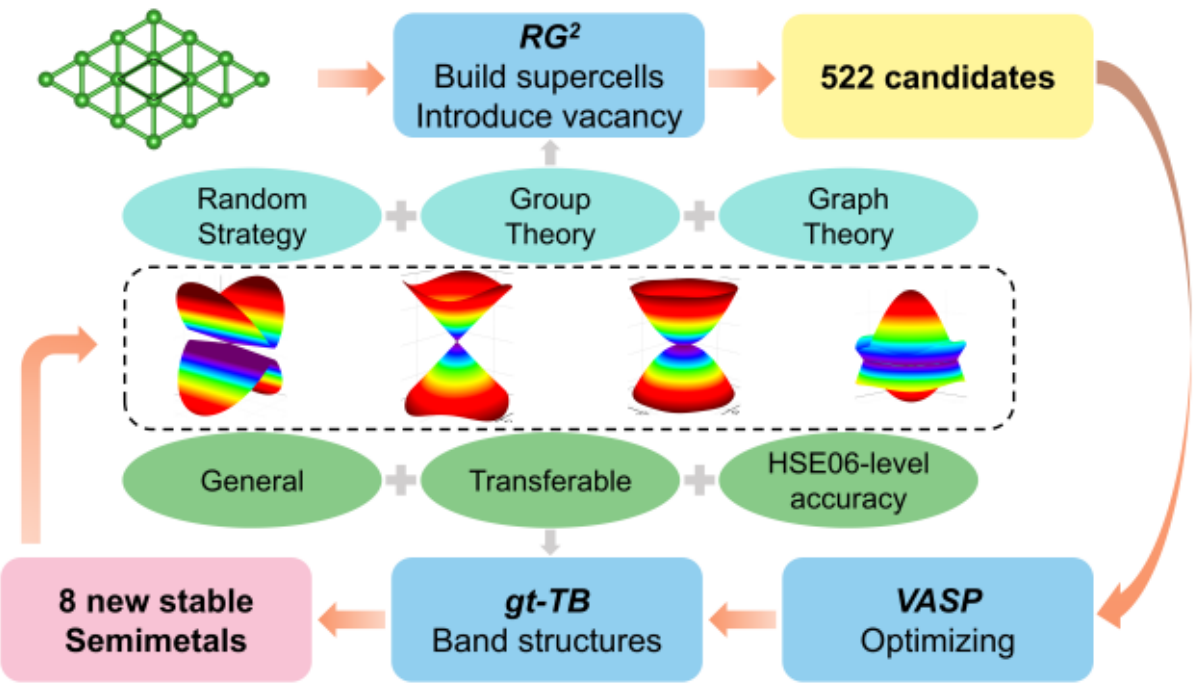

TOC Graphic

Adjacent to carbon in the periodic table, boron exhibits more complex bonding properties due to its electron deficiency nature, so a wide variety of allotropes can be formed, including clusters[1], fullerenes[2], nanotubes[3-4], two-dimensional sheets[5] and three-dimensional structures[6]. Among all boron nanostructures, two-dimensional sheets, namely borophene, have been extensively studied due to their interesting properties such as anisotropy[7], high mechanical strength, flexibility[8], and superconductivity[9-10]. Thousands of borophene configurations were predicted computationally via various structural search techniques, such as cluster expansion (CE) method[11-12], particle swarm optimization (PSO) global algorithm[13-14]. Experimentally, borophene cannot be obtained by mechanical exfoliation due to the lack of natural layered bulk boron. Instead, molecular beam epitaxy[7, 15-18], chemical vapor deposition[19], and sonochemical exfoliation[20] have been successfully employed to prepare various borophene phases. Notably, the large-scale growth of high-quality borophene and its transfer between different substrates have been achieved[21]. Recent studies have further demonstrated the synthesis of bilayer and few-layer borophenes[22-23], extending beyond the single-atomic-layer limit. This expansive allotropy demonstrates the potential for discovering borophene configurations with novel physical properties.

Over recent decades, topological materials have undergone rapid development, among which topological semimetals (TSMs) have garnered considerable interest as a prominent class due to their distinctive physical properties, such as chiral anomaly[24] and Fermi arcs[25]. TSMs can be classified into different categories based on the features of the band crossings, including nodal dimension (such as zero-dimensional nodal points, one-dimensional nodal lines, and two-dimensional nodal surfaces), band degeneracy (such as Weyl and Dirac), band slope (such as type I and II) and dispersion order (such as linear, quadratic, and cubic)[26]. Owing to its exceptional structural polymorphism, borophene has become a promising platform for discovering novel topological semimetals. The prediction in 2014 of buckled *Pmmn* borophene with anisotropic Dirac cones marked the emergence of the third elemental system hosting massless Dirac fermions, following graphene and silicene[27]. Subsequent theoretical studies have uncovered a variety of Dirac states in other borophene allotropes. For instance, monolayer hr-sB was predicted to host both Dirac nodal lines and tilted semi-Dirac cones near the Fermi level ($E_F$). Dirac cones at $E_F$ have been identified in bilayer P6/mmm[28] and planar $\chi$-$h_1$ [29] borophenes, while Dirac nodal lines at $E_F$ were predicted in bilayer $P\bar{6}$-boron[30], planar $\chi_{2/9}$[31] phases. Notably, the *Pmmn*, *P*6/*mmm*, $\chi$-$h_1$, $P\bar{6}$-boron and $\chi_{2/9}$ structures exhibit no extraneous bands at $E_F$, rendering them ideal semimetallic systems. Experimentally, Dirac cones in $\beta_{12}$ and $\chi_3$ structures have been confirmed via high-resolution angle-resolved photoemission spectroscopy (ARPES)[32-33].

Despite these advances, the majority of reported borophenes display metallic behavior, with ideal semimetallic phases remaining scarce. This scarcity raises two fundamental questions: (1) Are there more planar borophenes hosting intrinsic Dirac fermions near the Fermi level? (2) What other types of Dirac states might exist in borophene systems? Addressing these questions requires efficient exploration of the vast borophene configuration space, which in turn highlights a critical methodological challenge: how to accurately and efficiently characterize electronic properties using high-throughput computational strategies. In this work, we performed a high-throughput screening of planar semimetallic borophenes using a random strategy based on graph and group theory (RG$^2$) combined with first-principles calculations and transferable tight-binding (gt-TB) model. The workflow diagram is presented in FIG. 1. A group of TB parameters with excellent transferability for planar borophenes has been obtained, enabling efficient electronic band structure calculations for arbitrary configurations. Using the TB parameters, we identified eight novel borophenes hosting diverse Dirac states, including Type-I and Type-III Dirac cones, quadratic nodal points and Dirac nodal lines. Symmetry analyses confirm the topological protection of Dirac states. Their stabilities are confirmed by the phonon-spectrum and *ab initio* molecular dynamics (AIMD) simulation.

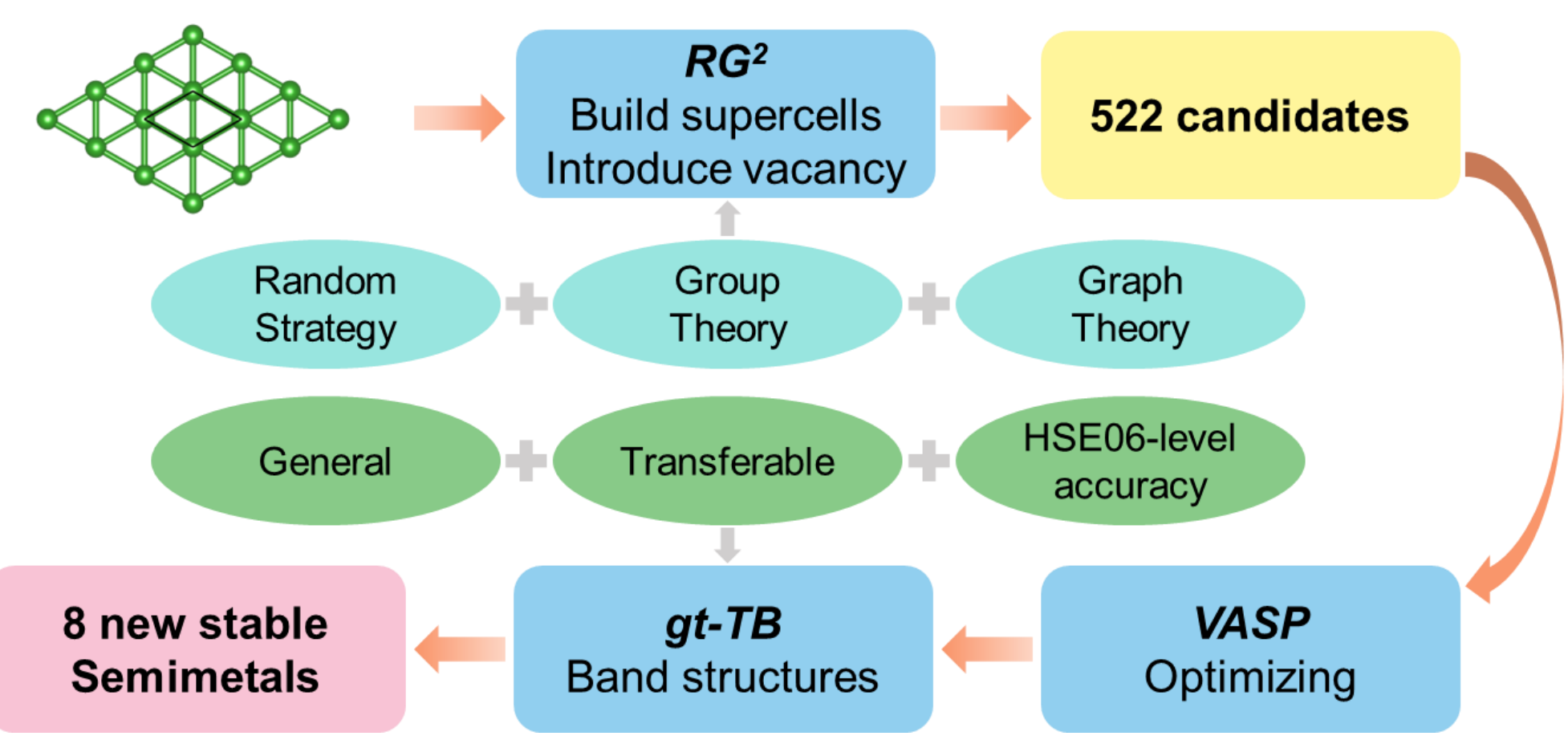

FIG. 1. Workflow diagram for high-throughput discovery of semimetallic borophenes

Planar borophenes can be modeled as periodically vacated triangular lattices, structurally characterized by the vacancy density $\eta$, which is defined as the ratio of vacancy sites to the number of atomic sites in the pristine triangular lattice[5]. Therefore, supercells were constructed based on the planar triangular $\delta_3$-borophene lattice, followed by the introduction of randomized vacancies. Symmetry-equivalent configurations were subsequently removed using group- and graph-theory algorithms implemented in the RG2 code[34]. To systematically characterize their structural configurations, all borophenes are labeled following the "sn-in-tn-rs" nomenclature based on their space group number (sn), inequivalent atom numbers (in), total atom numbers (tn), and ring sequences (rs).

The electronic structure calculations were performed using the Vienna ab initio simulation package (VASP). In the calculations, the projector augmented wave method and the Perdew−Burke−Ernzerhof generalized gradient approximation (GGA-PBE)[35] were used. The cutoff energy of plane wave was set to 480 eV. The convergence criteria for the total energy and forces acting on the boron atoms were set to $10^{-4}$ eV and 0.01 eV/Å, respectively. A vacuum layer of 15 Å was introduced to eliminate spurious interactions between adjacent periodic images. The Γ-centered k-point meshes with a uniform density of $2\pi \times 0.025$ Å$^{-1}$ generated by VASPKIT[36] were employed for all structures. In addition, the Heyd-Scuseria-Ernzerhof screened hybrid functional (HSE06)[37] was used to optimize the TB-parameters and to further validate the band structures of the newly identified topological semimetals.

To explore the dynamical stability of borophenes, phonon dispersions were calculated using the Phonopy package[38] interfaced with VASP, where the stringent convergence criteria for energy ($10^{-8}$ eV) and force ($10^{-3}$ eV/Å) were employed to ensure computational accuracy. To evaluate the thermodynamic stability, *ab initio* molecular dynamics (AIMD) simulations were performed with the temperature controlled at 500 K by a Nosé-Hoover thermostat[39]. A total simulation time of 5 ps with a time step of 1 fs was employed. To analyze the energy bands and their connectivity, the irreducible representations (IRs) are identified using the Irvsp package[40].

Since standard density functional theory (DFT) calculations using generalized gradient approximation (GGA) functionals tend to underestimate band gaps, the hybrid functional HSE06 is widely employed for more accurate electronic-structure calculations. However, HSE06 calculations are computationally expensive, rendering them impractical for large systems or high-throughput studies involving numerous structures. The Slater-Koster (SK)[41] based general and transferable tight-binding (gt-TB) method, which is both efficient and accurate, has been successfully applied to predict the electronic properties of silicon, carbon, and bismuth isomers[42-47]. Given this success, we constructed a gt-TB model for borophene structures to overcome this computational bottleneck. In this TB model, the corresponding Hamiltonian and overlap matrix are expressed as:

$$H = \sum_{i \neq i'; l \neq l'} t_{il,i'l'} (c_{il}^{+} c_{i'l'} + h.c.) + {}_{l} \sum_{il} c_{il}^{+} c_{il}$$

$$S = \sum_{i \neq i'; l \neq l'} s_{il,i'l'} (c_{il}^{+} c_{i'l'} + h.c.) + \sum_{il} c_{il}^{+} c_{il}$$

Where $i$ denote atomic site indices, $l$ denote orbital indices, and $c_{il}^{+}$ ($c_{il}$) represent the creation (annihilation) operator. $\varepsilon_l$ is the on-site energy of the orbital $l$, while $t_{il,i'l'}$ and $s_{il,i'l'}$ denote the hopping and overlap integrals, respectively. These integrals decrease rapidly with increasing interatomic distance and can be written as:

$$t_{il,i'l',\mu} = V_{ll'\mu} e^{q_1 \times (1 - d_{ii'}/d_0)}; \; s_{il,i'l',\mu} = S_{ll'\mu} e^{q_2 \times (1 - d_{ii'}/d_0)}$$

where $d_{ii'}$ is the distance between the $i$-th and $i'$-th atoms, and $d_0 = 1.68$ Å is taken as the reference B − B

bond length in $\delta_3$-borophene. The parameters $V_{ll'\mu}$ and $S_{ll'\mu}$ represent the reference values of the hopping and overlap integrals at $d_{ii'} = d_0$, while $q_1$ and $q_2$ control the decay rate of the respective integrals. All integrals are set to zero when the interatomic distance $d_{ii'}$ exceeds the cutoff distance $d_{cut} = 10$ Å . In this TB model, the complex and discrete integrals associated with different neighbors in different systems are unified through a single exponential decay formalism, thereby ensuring the transferability of the model.

To achieve HSE06-level accuracy within a computationally efficient framework, we first calculated the electronic structures of six representative borophene isomers using the hybrid HSE06 functional. The selection includes four experimentally synthesized phases (metal α, $\delta_3$, $\beta_{12}$ and $\chi_3$) and two theoretically predicted structures (semimetal $\chi$-$h_1$ and semiconductor $\beta_1^s$), as shown in FIG S1(a). The TB parameters were subsequently optimized by fitting to the HSE06 band structures using the Simplex algorithm[48], with the standard deviation between the TB and HSE06 band energies taken as the objective function. To further assess the transferability of the parameter set, additional HSE06 calculations were performed for six previously reported borophene allotropes (FIG. S1(b)).

Table 1 Optimized tight-binding parameters for borophene. The $V$ and $E_{onsite}$ parameters are in eV, and the $S$ parameters are dimensionless.

| | | ***ssσ*** | ***spσ*** | ***ppσ*** | ***ppπ*** | ***sdσ*** | ***pdσ*** | ***pdπ*** | ***ddσ*** | ***ddπ*** | ***ddδ*** |
|---|---|---|---|---|---|---|---|---|---|---|---|
| ***t*** | $V$ | -4.805 | 4.989 | 5.611 | -2.644 | -1.387 | -0.291 | 1.576 | -0.836 | -0.339 | -0.977 |
| | $q_1$ | 3.654 | 3.055 | 2.020 | 3.846 | 0.893 | 1.014 | 2.223 | 0.857 | 9.977 | 1.243 |
| ***s*** | $S$ | 0.114 | -0.094 | -0.123 | 0.064 | 0.008 | -0.095 | -0.047 | 0.284 | 0.325 | 0.157 |
| | $q_2$ | 6.384 | 8.101 | 1.156 | 8.872 | 0.402 | 2.076 | 11.690 | 1.140 | 1.417 | 12.181 |
| $\boldsymbol{E_{onsite}}$ | $E_s$=-7.863 | | $E_p$=-2.609 | | $E_d$=1.935 | | | | | | |

The selection of appropriate basis functions is critical for constructing an accurate and transferable TB model. Previous TB descriptions of borophenes based on either $p$-only or $sp^3$ basis sets can reasonably reproduce the electronic structures of individual allotropes[49-51]. However, earlier studies on group III–V systems have demonstrated that $d$ orbitals play a critical role, particularly in describing unoccupied electronic states[42, 52-55]. To systematically evaluate the effect of basis selection, we compared TB models constructed using $sp^3$ and $sp^3d^5$ basis sets. The optimized TB parameters with $sp^3$ basis are summarized in Table S 1, and the corresponding TB band structures of the borophenes for parameter fitting and validation are shown in FIG S 2 and FIG S 3, respectively. We find that the $sp^3$-basis model provides only a rough approximation of the electronic structures and exhibits limited transferability across different allotropes. In contrast, the $sp^3d^5$-basis model reproduces the HSE06 band structures with remarkable accuracy over an energy window extending up to 4 eV above the Fermi level, as shown in FIG. 2. Its transferability is further validated by excellent agreement with HSE06 results for six additional borophene systems (FIG S 4). The final optimized $sp^3d^5$-basis parameters are listed in Table 1. Collectively, these results demonstrate that the proposed $sp^3d^5$ TB model is transferable and reliable for electronic band structure evaluations of arbitrary planar borophenes.

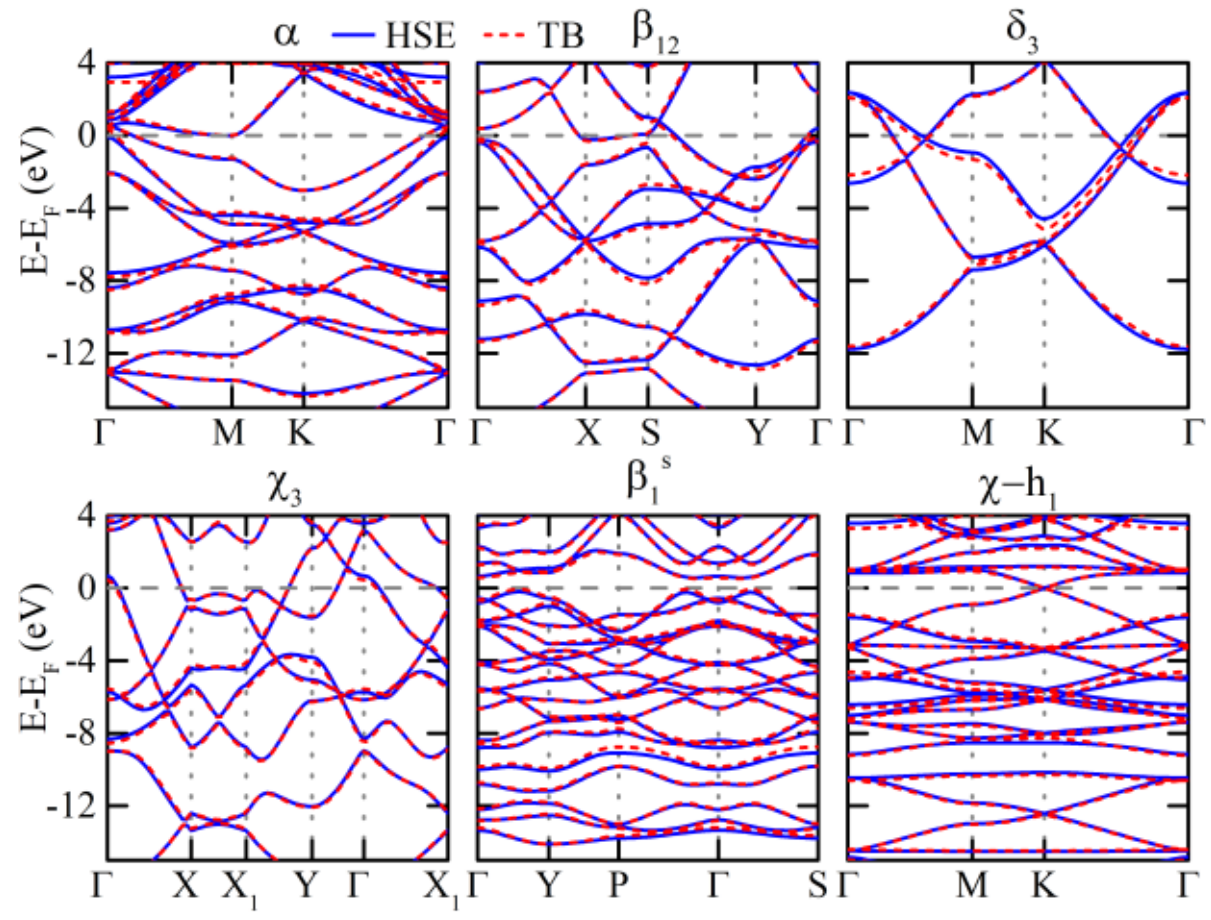

FIG. 2. Comparison between HSE06 and $sp^3d^5$-TB band structures for the six borophene allotropes used in parameter fitting.

By introducing vacancies into $\delta_3$ supercells via RG[2] code, 522 inequivalent borophene structures ($0 < \eta < 0.528$) were obtained. Notably, this dataset naturally recovers many previously proposed borophene isomers, including α sheet, β sheet[5], γ-sheet ($\beta_{12}$)[2], $\beta_4$(1/8) ,4/27, $\beta_5$(2/15) [11],1/12($\alpha_5$), struc-e($\delta_5$), struc-1/4, 1/5($\chi_3$), struc-1/10[14], and the topological nodal-line semimetal B10[18], $\chi_{2/9}$[31], semimetallic borophene χ-$h_1$, χ-$h_2$[29]. The successful reproduction of these known phases confirms the reliability and completeness of the structural search strategy. We note that the hole density η of the $\chi_{2/9}$ phase should be 1/3 rather than 2/9.

Following structural relaxation with VASP, TB band structures for all configurations were rapidly evaluated using the optimized parameter set. From this large-scale screening, eight previously unreported borophene allotropes exhibiting distinct band crossings near the Fermi level were identified. Subsequent HSE06 calculations confirmed both their semimetallic nature and the transferability of the TB model. These newly discovered allotropes host a rich variety of topological fermionic states, including type-I/III Dirac cones (65-4-24-r68, 65-4-24-rx, 47-5-14, and 189-5-23), quadratic band-crossing points (174-6-16, 174-6-18), and Dirac nodal loops (51-10-36 and 175-4-24). Structural parameters for all newly identified borophenes are tabulated in Table S2.

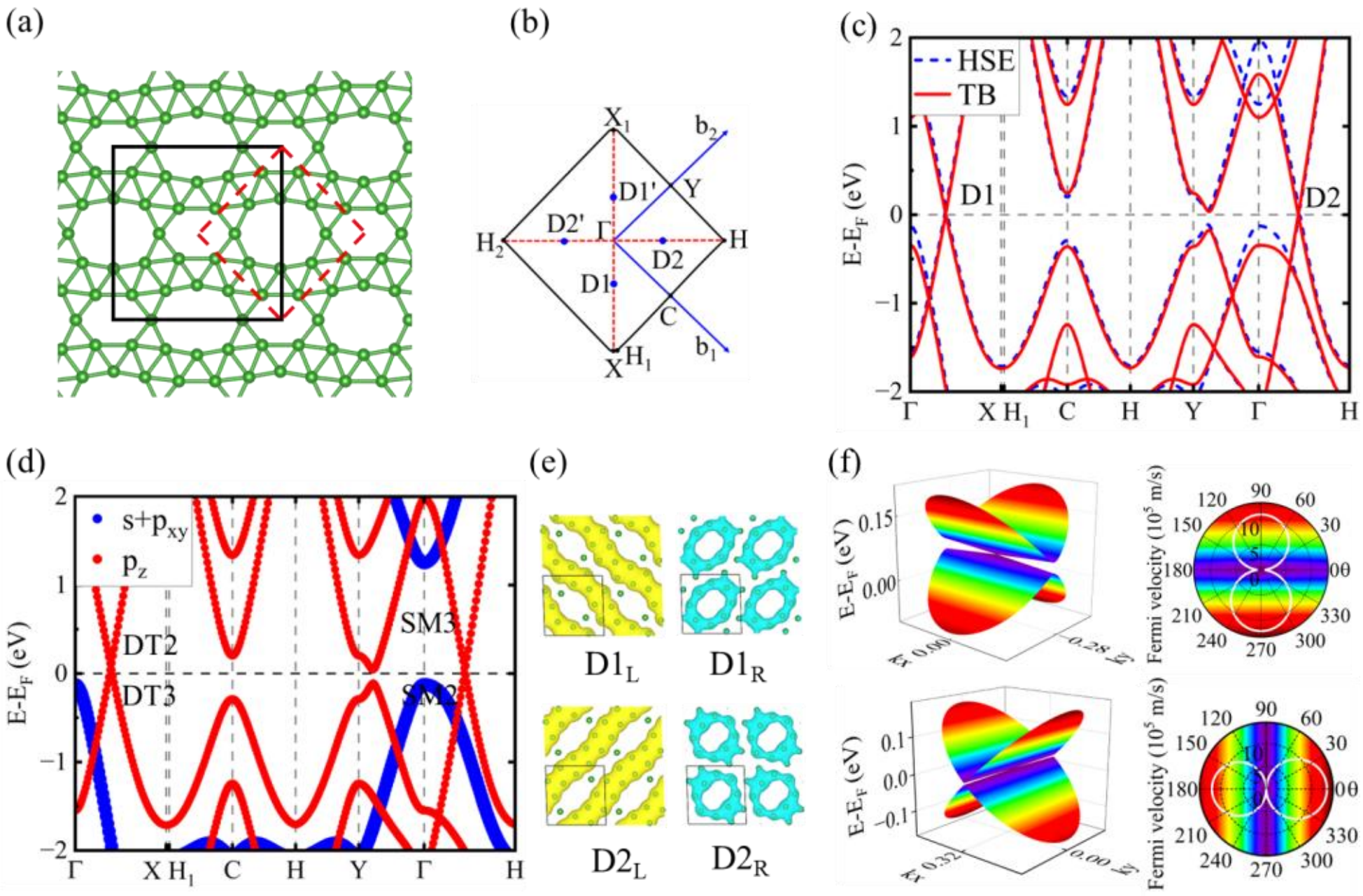

FIG. 3. (a) Crystal structure of 65-4-24-r68, black lines indicate the conventional cell, red dashed lines indicate the primitive cell. (b) The positions of the Dirac points in the Brillouin zone (BZ). (c) Band structures by HSE06 and TB. (d) Orbital-resolved band structure, where the size of dots is proportional to the weight contributed by different orbitals. (e) The band-decomposed charge densities. $D_L$ and $D_R$ denote the left and right side of the Dirac point, respectively. (f) Left: the 3D band structures in the vicinity of nodes D1 and D2. Right: The direction-dependent Fermi velocities calculated based on HSE06 results (white line) of D1 and D2.

FIG. 3(a) presents the optimized crystal structure of 65-4-24-r68, which crystallizes in the orthorhombic *Cmmm* space group and contains 24 boron atoms within the conventional unit cell. The structure possesses vertical mirror planes and $C_2$ rotational symmetries along the *x*-, *y*- and *z*-directions. The corresponding TB and HSE06 band structures along high-symmetry paths are shown in FIG. 3(c), revealing two Dirac points: D1 at (0.196, -0.196, 0) along Γ–X path and D2 at (0.217, 0.217, 0) along Γ–$H_1$ path. Symmetry operations generate two additional equivalent Dirac points in the Brillouin zone [Fig. 3(b)].

To explore the origin of the Dirac cones, the orbital-resolved band structures are calculated as shown in FIG. 3(d). It is obvious that both Dirac cones are dominated by the $p_z$ orbitals. To verify the band crossings, we calculated the decomposed charge densities near the Dirac points [FIG. 3(e)]. The charge density inversion across each Dirac point confirms band crossing formation. Moreover, D1 and D2 exhibit similar charge distributions—both characterized by ring patterns and stripe patterns.

The HSE06-based 3D band structures of 65-4-24-r68 near D1 and D2, shown in FIG. 3(f) reveal strongly distorted and highly anisotropic Dirac cones. To quantify this anisotropy, we evaluated the direction-dependent Fermi velocity according to $v_f = E(k)/\hbar|k|$. The Fermi velocities of D1 vary from $8.758\times 10^4\ m/s$ (θ=0°) to $1.214\times 10^6\ m/s$(D1–X), whereas that of D2 spans from $1.102\times 10^4\ m/s$ (θ=90°) to $1.335\times 10^6\ m/s$(D2–H). The anisotropy of the Dirac cone [A(D)] can be defined as [A(D)] = $(v_{max}(\theta) - v_{min}(\theta))/(v_{max}(\theta) + v_{min}(\theta))$, where $v_{max}(\theta)$ and $v_{min}(\theta)$ are the maximum and minimum direction-dependent Fermi velocities. Based on the data in FIG. 3, we can easily obtain large anisotropies of A(D1) = 86.54 % and A(D2) = 98.36 %. Such pronounced intrinsic anisotropy, achieved without requiring external manipulation, highlights the potential of this allotrope for highly directional quantum transport and anisotropic electronic applications.

To investigate the formation mechanism of the Dirac points, the group symmetry and parity of bands were analyzed using the Irvsp code and the Bilbao Crystallographic Server[56]. The IRs of the crossing bands at D1 are identified as DT2 and DT3, which belong to the two opposite eigen-wave-functions of $M_x$ mirror reflection and $C_{2y}$ rotation. Similarly, the IRs of the crossing bands at D2 are confirmed as SM2 and SM3, which belong to the two opposite eigen-wave-functions of $M_y$ mirror reflection and $C_{2x}$ rotation. Therefore, the band inversion in 65-4-24-r68 borophene should be caused and protected by out-of-plane mirror and $C_2$ rotation symmetry.

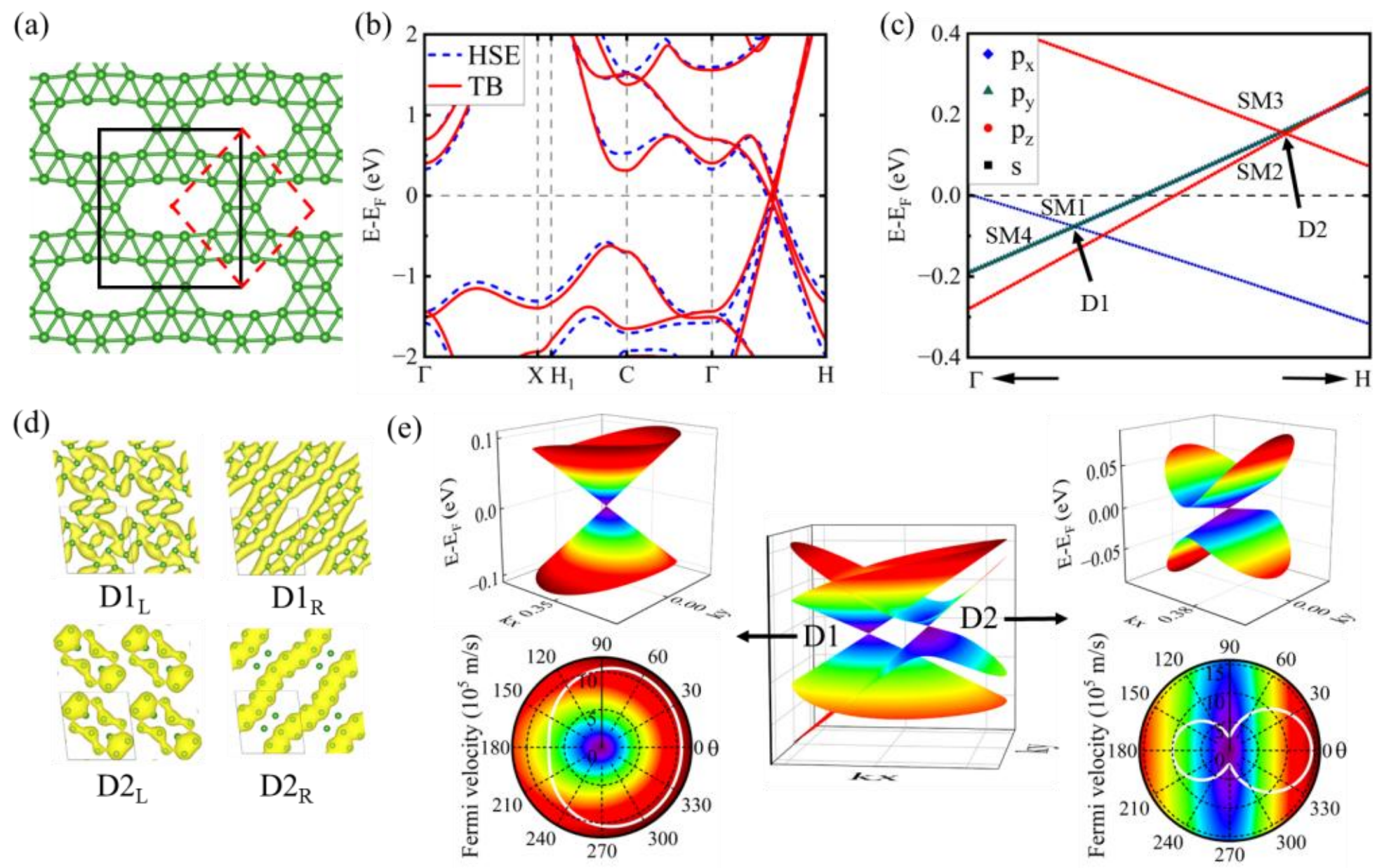

FIG. 4. (a) Crystal structure of 65-4-24-rx, black lines and red dashed lines indicate the conventional cell and primitive cell, respectively. (b) Band structures by HSE06 and TB. (c) Orbital-resolved band structure, where the size of dots is proportional to the weight contributed by different orbitals. (d) The band-decomposed charge densities. $D_L$ and $D_R$ denote the left and right side of the Dirac point, respectively. (e) The 3D band structures and the direction-dependent Fermi velocities of D1 and D2.

Another structure 65-4-24-rx in space group *Cmmm* exhibits four band crossings near the Fermi level along the Γ–H path [FIG. 4(b)]. To characterize these crossings, we computed the orbital-projected band structure, charge density distribution, and 3D band dispersion. The 3D band reveals that two of these crossings correspond to isolated Dirac cones (D1 and D2), while the other two belong to Dirac nodal lines generated by cone nesting. Based on the orbital-projected band structure and partial charge density analysis: D1 originates from the crossing between $p_x$- and $p_y$-dominated bands, exhibiting weak anisotropy A(D1) = 21.07 %, with a maximum Fermi velocity of $1.083 \times 10^6$ *m/s*. In contrast, D2 arises from two $p_z$-dominated bands, showing strong anisotropy A(D2) = 70.46%, reaching a maximum Fermi velocity of $1.432 \times 10^6$ m/s. Symmetry analysis confirms that both Dirac cones are protected by either $C_{2x}$ rotation or $M_y$ mirror symmetry.

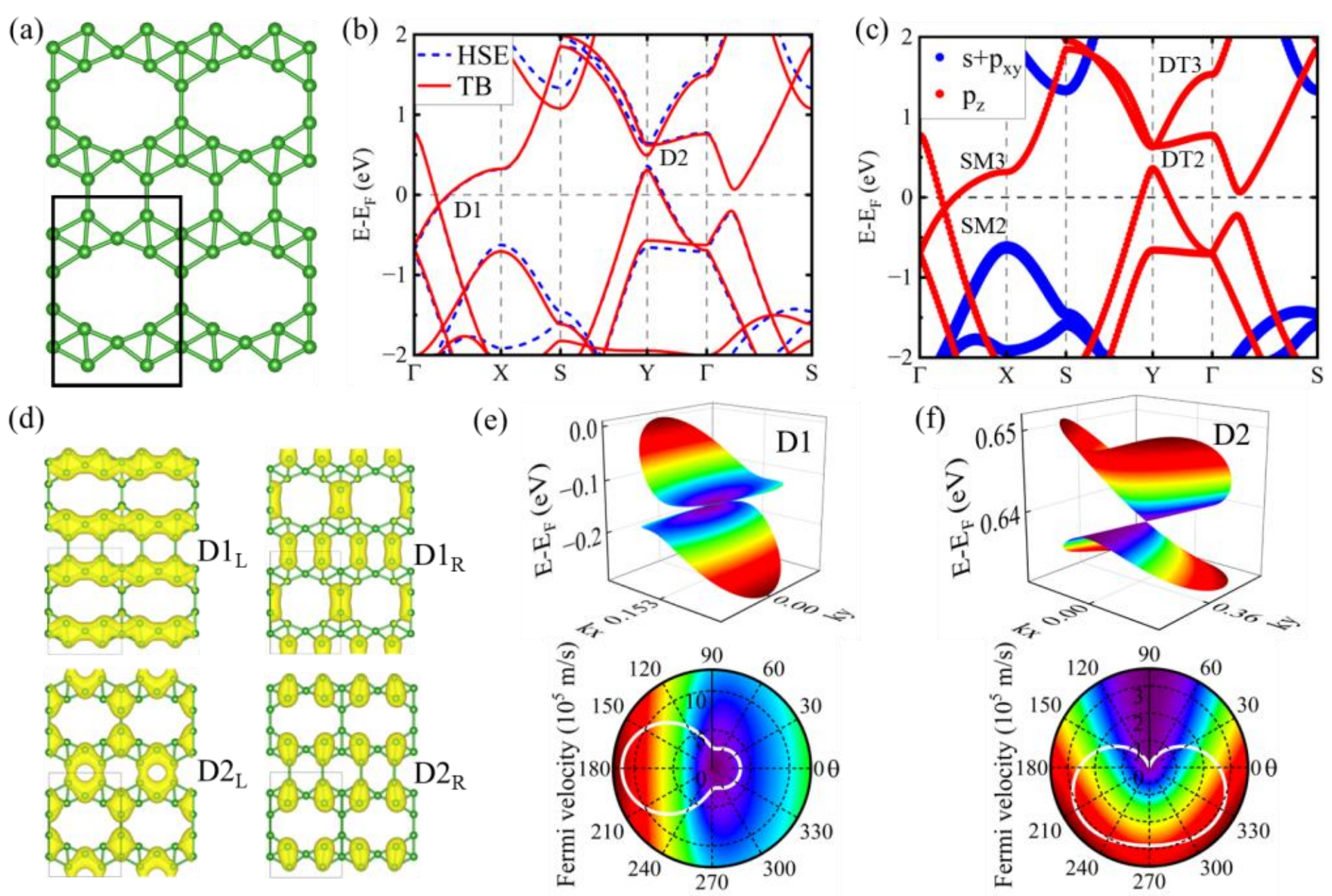

FIG. 5. (a) Crystal structure of 47-5-14. (b) Band structures of 47-5-14 by HSE06 and TB. (c) Orbital-projected band structure. (d) The band-decomposed charge densities. $D_L$ and $D_R$ denote the left and right side of the Dirac point, respectively. (e)(f) The 3D band structures and the direction-dependent Fermi velocities (white line) for D1 and D2.

As evidenced in FIG. 5 and 6, the borophenes 47-5-14 and 189-5-23 host Dirac cones near $E_F$ with band crossings formed by $p_z$ orbitals, which are also confirmed through partial charge densities in FIG. 5(d) and FIG. 6(d). The 3D band structures in FIG. 5(e)(f) reveal strong anisotropy of Dirac cones in 47-5-14. D1 exhibits a type-I configuration with A(D1) = 64.73%, while D2 manifests a type-III characteristic with an exceptionally large anisotropy of 94.45%. By contrast, 189-5-23 hosts nearly isotropic graphene-like Dirac cones. Notably, borophene 189-5-23 also exhibits relatively flat bands originating from its coloring-triangle (CT) lattice structure, which is topologically equivalent to the Kagome lattice[57], as shown in FIG. 6(a). Symmetry analysis establishes topological protection via $C_2$ symmetries for both systems.

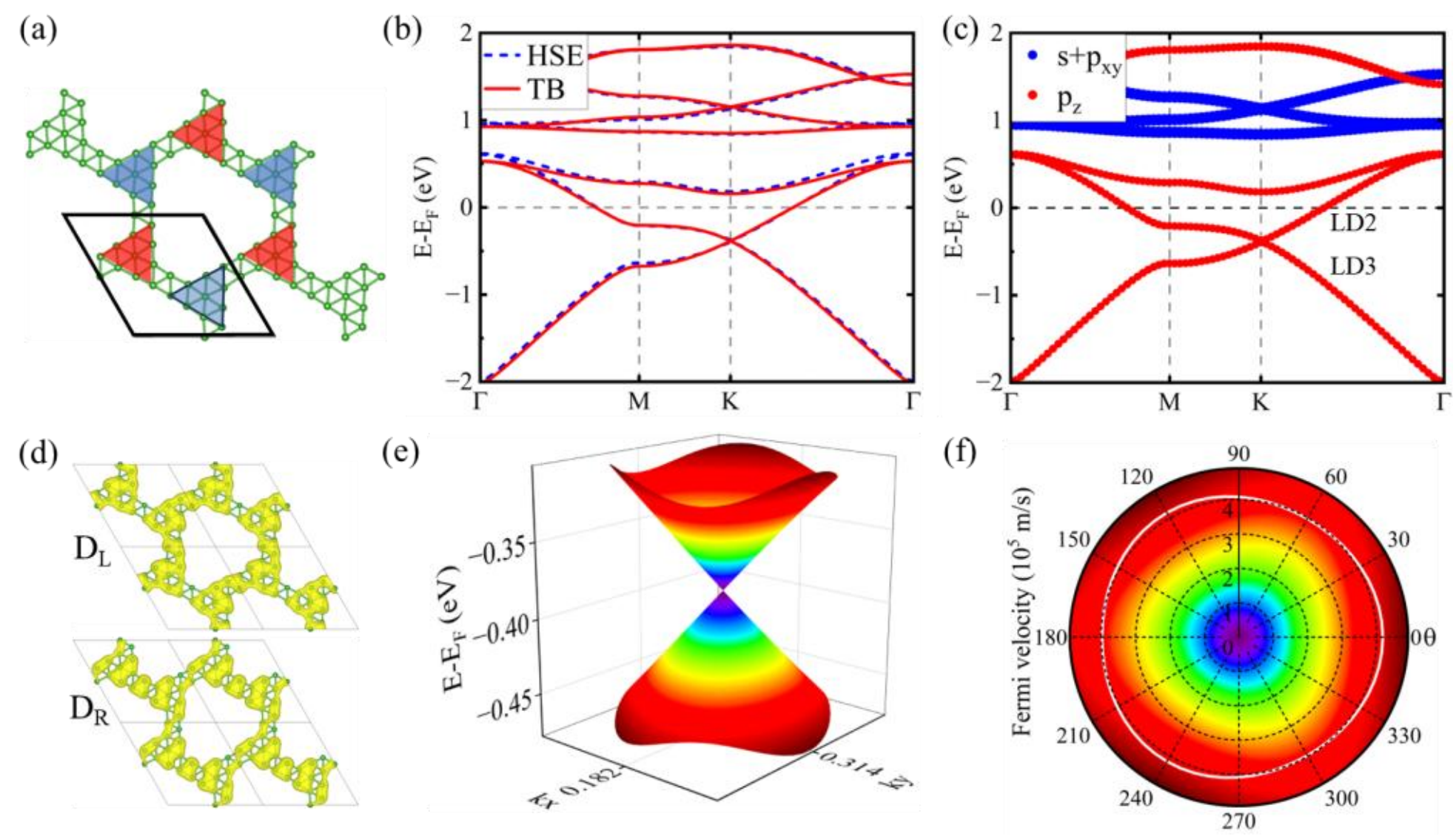

FIG. 6. (a) Crystal structure of 189-5-23. (b) Band structures of 189-5-23 by HSE06 and TB. (c) Orbital-resolved band structure. (d) The band-decomposed charge densities. $D_L$ and $D_R$ denote the left and right side of the Dirac point, respectively. (e) The 3D band structures in the vicinity of the Dirac point. (f)The direction-dependent Fermi velocity (white line).

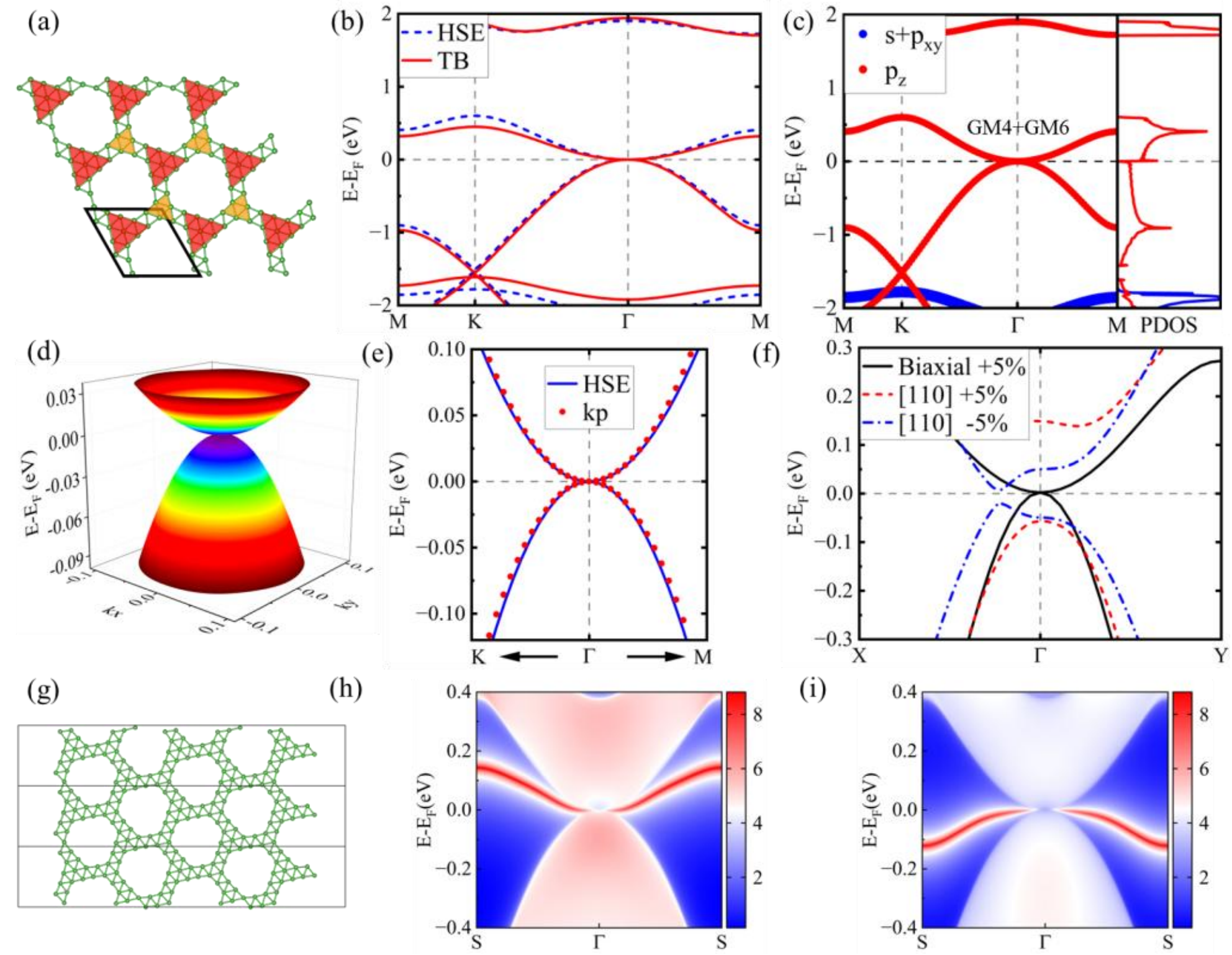

FIG. 7. (a) Crystal structure of 174-6-16, the larger triangles are B10 clusters and smaller triangles are B6 clusters. (b) Band structures of 174-6-16 by HSE06 and TB. (c) Orbital-projected band structure and PDOS. (d) The 3D band structures in the vicinity of the Γ point. (e) Enlarged view of the band dispersion by HSE06 (blue line) and *k·p* Hamiltonian (red points) around Γ. (f) Band structures of borophene under biaxial +5% strain and uniaxial ±5% strain along [110]. (g) 174-6-16 nanoribbon. (h) and (i) Edge states of the nanoribbon.

The borophene allotrope 174-6-16 is composed of interconnected B6 and B10 clusters [Fig. 7(a)] and crystallizes in the $P\bar{6}$ space group. The structure preserves $C_{3z}$ rotational symmetry and out-of-plane mirror symmetry $M_z$, while lacking inversion symmetry. As shown in Fig. 7(b), the highest valence band (HVB) and lowest conduction band (LCB) touch at the Γpoint, and both bands exhibit quadratic rather than linear dispersion. Orbital-projected band structures and projected density of states analyses confirm the dominant contribution of $p_z$ orbitals to the crossing bands.

Unlike Dirac systems with linear dispersion, quadratic band crossings produce a significantly enhanced density of states near the Fermi level, which can potentially stabilize interaction-driven correlated quantum phases[58-59]. The 3D band structure around the Γ point (FIG. 7d) clearly demonstrates isotropic quadratic dispersion in all momentum directions, identifying 174-6-16 as a quadratic contact-point (QCP) semimetal.

Symmetry analysis reveals that the touching bands belong to the doubly degenerate irreducible representation "GM4+GM6" at Γ. Along the Γ-K and Γ-M directions, the corresponding representations reduce to LD2 and SM2, respectively. The generators of the Γ-little group are {C3z|0, 0, 0} and {Mz|0, 0, 0}. Based on these states, a low-energy effective Hamiltonian at Γ was constructed automatically using package VASP2KP[60]. To the second order, the *k·p* Hamiltonian is expressed in the following form:

$$H^{kp} = \begin{pmatrix} D_1 & D_2 \\ \dagger & D_1 \end{pmatrix},$$

$$D1 = c_1\left(k_x^2 + k_y^2\right),$$

$$D_2 = c_2(k_x + ik_y)^2 - ic_3(k_x + ik_y)^2,$$
$$c_1 = -0.59,\ c_2 = -4.24,\ c_3 = -2.728$$

The energy is analytically obtained as

$$E = (-0.59 \pm 5.04)(k_x^2 + k_y^2)$$

Clearly, only quadratic terms persist in this expansion, directly reflecting the restrictions imposed by $C_{3z}$ and $M_z$ symmetries. A biaxial tensile strain of 5% preserves crystal symmetry and maintains the quadratic nodal point, whereas uniaxial strain or compression along the [110] direction break the $C_3$ rotational symmetry and lift the band degeneracy.

To compute the edge spectrum, we consider a semi-infinite nanoribbon shown in FIG. 7(g). The calculations are carried out via WannierTools[61] package using A tight-binding Hamiltonian obtained by Wannier90[62]. The resulting edge states emerge directly from the quadratic nodal point in FIG. 7(h) and 7(i), further supporting its nontrivial topological character.

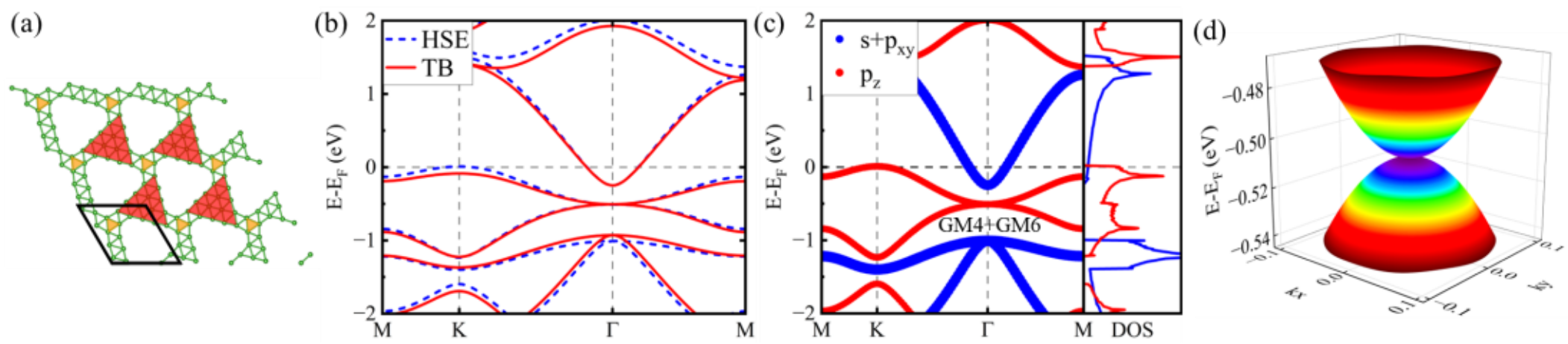

FIG. 8. (a) Crystal structure of 174-6-18. (b) Band structures of 174-6-18 by HSE06 and TB. (c) Projected band structure and PDOS. (d) The 3D band structures in the vicinity of the Γ point.

Motivated by the symmetry-protection mechanism identified above, we extended our screening to additional structures belonging to space group 174 and identified a second quadratic-contact semimetal, 174-6-18. Similar to 174-6-16, this allotrope hosts a symmetry-protected quadratic nodal point, although the crossing lies approximately 0.5 eV below the Fermi level as shown in FIG. 8. Importantly, both systems exhibit isolated nodal features with minimal interference from trivial bands, making them promising candidates for direct experimental observation via ARPES.

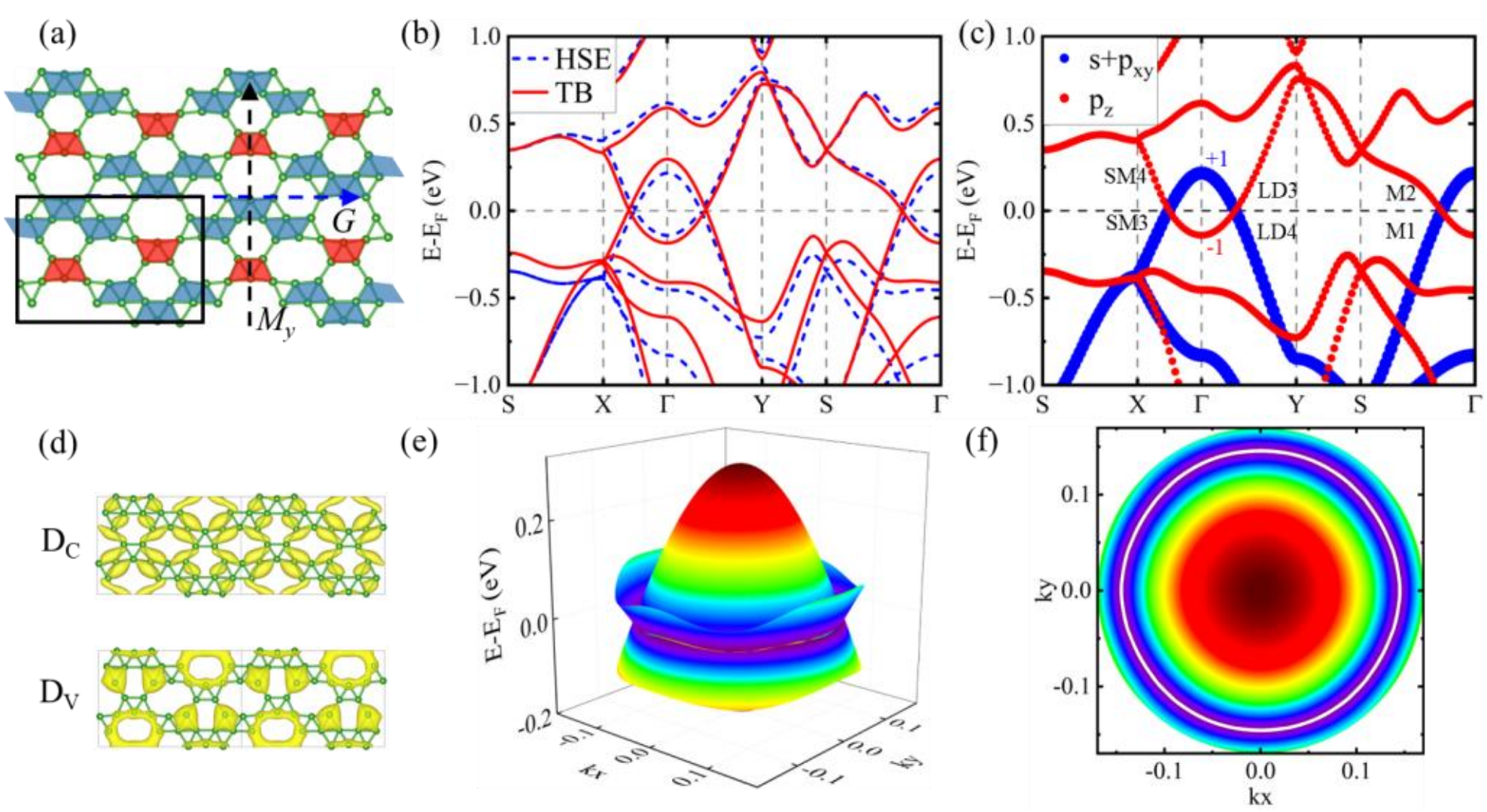

FIG. 9. (a) Crystal structure of 51-10-36. (b) Band structures of 51-10-36 by HSE06 and TB. (c) Orbital-resolved band structure. (d) The band-decomposed charge densities of Γ point. $D_C$ and $D_V$ denote the conduction band and valence band, respectively. (e) The 3D band structure around Γ point. (f) The energy difference of the LCB and HVB, where the white circle is the Dirac nodal line.

Borophene 51-10-36, belonging to space group *Pmam*, simultaneously hosts glide-mirror and mirror symmetries. The atomic structure is shown in FIG. 9(a), where the glide mirror plane and mirror plane are indicated by blue and black arrows, respectively. As shown in FIG. 9(b), its electronic structure exhibits multiple Dirac crossings along the Γ–X, Γ–Y, and Γ–S directions exactly at the Fermi level, indicating a zero-gap semimetallic state. Orbital-resolved band structures in FIG. 9(c) reveal that the HVB around Γ is dominated by the $p_z$ orbitals, while the LCB mainly originates from hybridized $s$, $p_x$ and $p_y$ orbitals, explaining the bands crossings. In addition, the partial charge densities of Γ point shown in FIG. 9(d) are fully consistent with the orbital picture.

To study the morphology of the band crossing, we plot the TB-based 3d band structure and map the energy difference between the HVB and LCB in FIG. 9(e) and 9(f). The distribution of band-crossing points, highlighted as a white line, reveals a closed nodal ring surrounding the Γ point, thereby confirming the Dirac nodal-line character of 51-10-36. The Fermi velocity of LCB(HVB) is $4.597 \times 10^5 m/s$ ($4.024 \times 10^5 m/s$) in the X − Γ direction, $4.895 \times 10^5 m/s$ ($3.382 \times 10^5 m/s$) in the Γ − Y direction and $4.989 \times 10^5 m/s$ ($3.108 \times 10^5 m/s$) in the S − Γ direction. These velocities approach the characteristic Fermi velocity range of graphene.

To elucidate the symmetry protection mechanism, we analyzed the group symmetries and parity eigenvalues. Our analysis reveals that the protection of the Dirac point is governed by distinct symmetry operations along different high-symmetry paths: it is protected by both $M_z$ and a glide symmetry along Γ–X, by $M_z$ and $M_y$ along Γ–Y, and by $M_z$ alone along Γ–S. Therefore, the nodal ring is protected by $M_z$ symmetry in the whole Brillouin zone. In contrast, the crossing bands of $\chi_{2/9}$ are dominated by the $p_z$ orbitals and protected by the out-of-plane mirror or $C_2$ symmetry, resulting a gapped nodal loop[31]. While in the bilayer case, the nodal lines of $P6/mmm$ and $P\bar{6}$ boron are formed by the hybrid states of $s$ orbital and $p$ orbitals[28, 30].

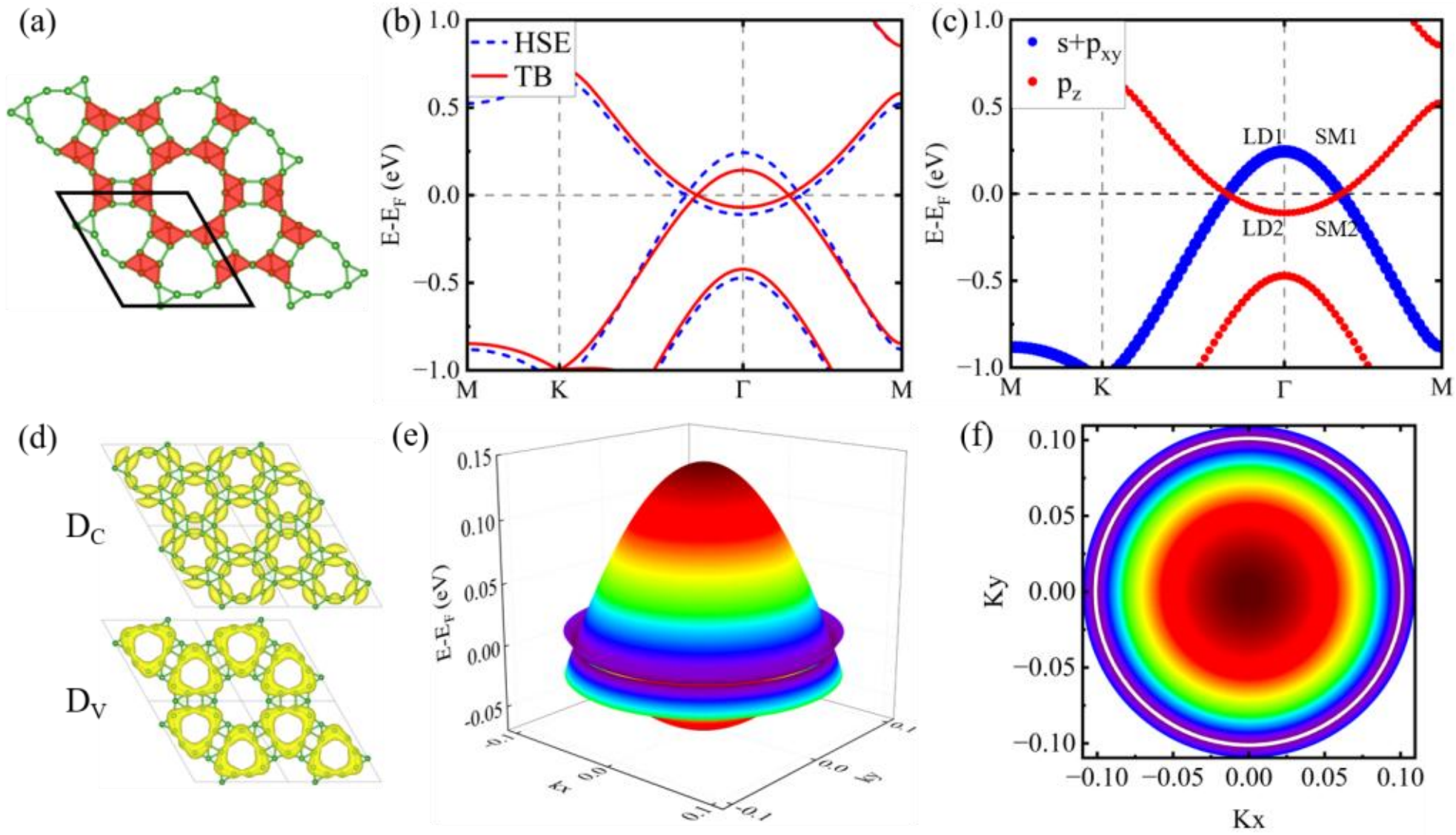

FIG. 10. (a) Crystal structure of 175-4-24. (b) Band structures of 175-4-24 by HSE06 and TB. (c) Orbital-resolved band structure. (d) The band-decomposed charge densities of Γ point. $D_C$ and $D_V$ denote the conduction band and valence band, respectively. (e) The 3D band structure around Γ point. (f) The energy difference between the LCB and HVB, where the white line is the Dirac nodal line.

As illustrated in FIG. 10(b), the band structures of 175-4-24 show two linear bands crossing on the K–Γ and Γ–M highly symmetric lines at the Fermi level, indicating a possible Dirac nodal line. The orbital-resolved band structure and the band-decomposed charge densities explain the bands crossing. The TB-based 3D band structure and contour mapping of the conduction–valence energy difference in FIG. 10(e)(f) unambiguously demonstrate the existence of a closed nodal line. The group symmetry analysis suggests that the nodal line is protected by the in-plane mirror symmetry, which is same as the 51-10-36. By calculating the slope of HVB and LCB, it can be found that the Fermi velocity of HVB is $2.6 \times 10^5\ m/s$ and the Fermi velocity of LCB is $5.4 \times 10^5\ m/s$, comparable to that of graphene.

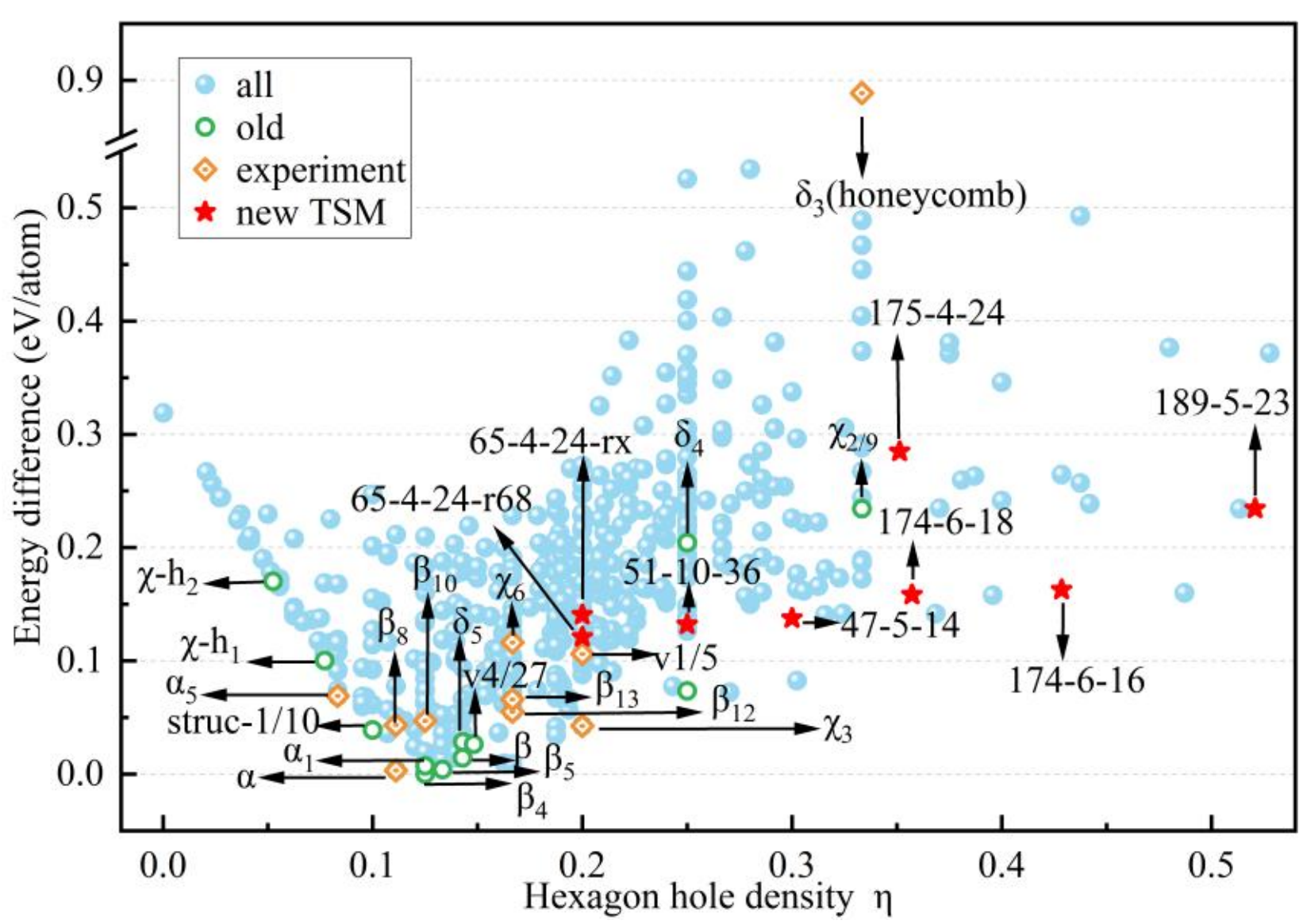

FIG. 11. Energy difference as a function of hexagonal vacancy density (η) for borophene allotropes.

To assess structural stability, we calculated the relative total energies of both the newly predicted allotropes and previously reported borophenes, taking the energetically most favorable $\beta_4$ phase as the reference in FIG. 11. It is evident that the lowest-energy configurations are distributed in the hole density $1/9 < \eta < 1/5$ interval, which is consistent with previous studies[11, 14]. The relative energies indicate that all newly identified structures are promising candidates. The energy of 175-4-24 and 189-5-23 are slightly higher than that of the Dirac nodal-line candidate $\chi_{2/9}$ but lower than that of the synthesized $\delta_3$ borophene. The other six isomers are energetically close to the experimentally synthesized $\chi_6$ phase, underscoring their potential for experimental realization. The phonon dispersions of the newly proposed borophenes are shown in FIG. S5, the absence of imaginary frequency strongly suggests they are dynamically stable. The AIMD simulations in FIG. S6 show that the atomic structures of the newly proposed borophenes remain intact and the total energy slightly around the equilibrium state energy during the entire simulation time, suggesting that all the structures are thermally stable at 500K.

It is interesting to note that the newly discovered semimetallic borophenes exhibit η ranging from 0.2 to 0.52, corresponding to structures featuring large holes with ring sizes including 3-, 4-, 6-, 8-, 9-, 12-, 15-, and 21-membered rings. These expanded ring configurations are accompanied by an increased proportion of four-coordinated atoms located at the hole edges, and the coordination number (CN) of boron atoms are summarized in Supplementary Table S2. Previous studies have shown that borophenes featuring higher vacancy density ($\eta > 20$ %) tend to form porous structures and exhibit greater energetic stability on metal or boride substrates[63-64]. Consequently, these borophene configurations might be synthesized by careful experimental designs.

In this study, we developed a transferable TB parameter set for planar borophenes that enables high-throughput electronic band structure calculations with HSE06-level accuracy. By combining this TB model with first-principles validation, we systematically screened 522 $RG^2$-generated candidates and identified eight previously unknown semimetallic borophenes hosting diverse Dirac states, including Type-I and Type-III Dirac cones, Dirac nodal lines, and the first predicted quadratic nodal points in borophene systems. Comprehensive symmetry analyses reveal that the Dirac cones are protected primarily by crystalline rotational symmetries, whereas the nodal loops originate from mirror-symmetry protection. The phonon spectrum and AIMD simulations further confirm the dynamical and thermal stabilities of all predicted structures. These findings significantly expand the family of two-dimensional boron allotropes and provide a material platform for exploring novel quantum phenomena arising from engineered Dirac states.

**Acknowledgements**

This work is supported by the National Natural Science Foundation of China (Grant No. 52372260), the Youth Science and Technology Talent Project of Hunan Province (Grant No. 2022RC1197), the Scientific Research Foundation of Education Bureau of Hunan Province (Grant Nos.20A503 and 23A0102), the Science Fund for Distinguished Young Scholars of Hunan Province of China (Grant Nos. 2021JJ10036 and 2024JJ2048), the Natural Science Foundation of Hunan Province (Grant No. 2022JJ30554) and the Postgraduate Scientific Research Innovation Project of Xiangtan University (Grant No. XDCX2025Y272). T.O. acknowledges the National Natural Science Foundation of China (Grant No. 52372260), the Youth Science and Technology Talent Project of Hunan Province (Grant No. 2022RC1197) and the Science Fund for Distinguished Young Scholars of Hunan Province of China (Grant No. 2024JJ2048). Y.S. acknowledges the Postgraduate Scientific Research Innovation Project of Xiangtan University (Grant Nos. XDCX2025Y272).

**Supporting Information Available:**

Borophene structures used for TB fitting and validation; optimized $sp^3$-basis TB parameters; comparisons of VASP and TB band structures for both $sp^3$ and $sp^3d^5$ models; structural parameters of the semimetallic borophenes; and phonon spectra together with AIMD simulation results demonstrating structural stability (DOC).

# Supplementary for " High-Throughput Discovery of Semimetallic Borophenes with Diverse Dirac States Via Transferable Tight-Binding Approach"

Yuke Song[1,2], Shifang Li[1,2], Tao Ouyang[1,2], Chao Tang[1,2], Jin Li[1,2*] and Chaoyu He[1,2*]

3. School of Physics and Optoelectronics, Xiangtan University, Hunan 411105, China
4. Hunan Provincial Key Laboratory of Computational Condensed Matter Physics and Quantum Materials Engineering, Hunan 411105, China

*Corresponding author. Email: lijin@xtu.edu.cn (Jin Li), hechaoyu@xtu.edu.cn (Chaoyu He)

**Contents:**

## 1. Borophene structures for fitting and validating the TB parameters.

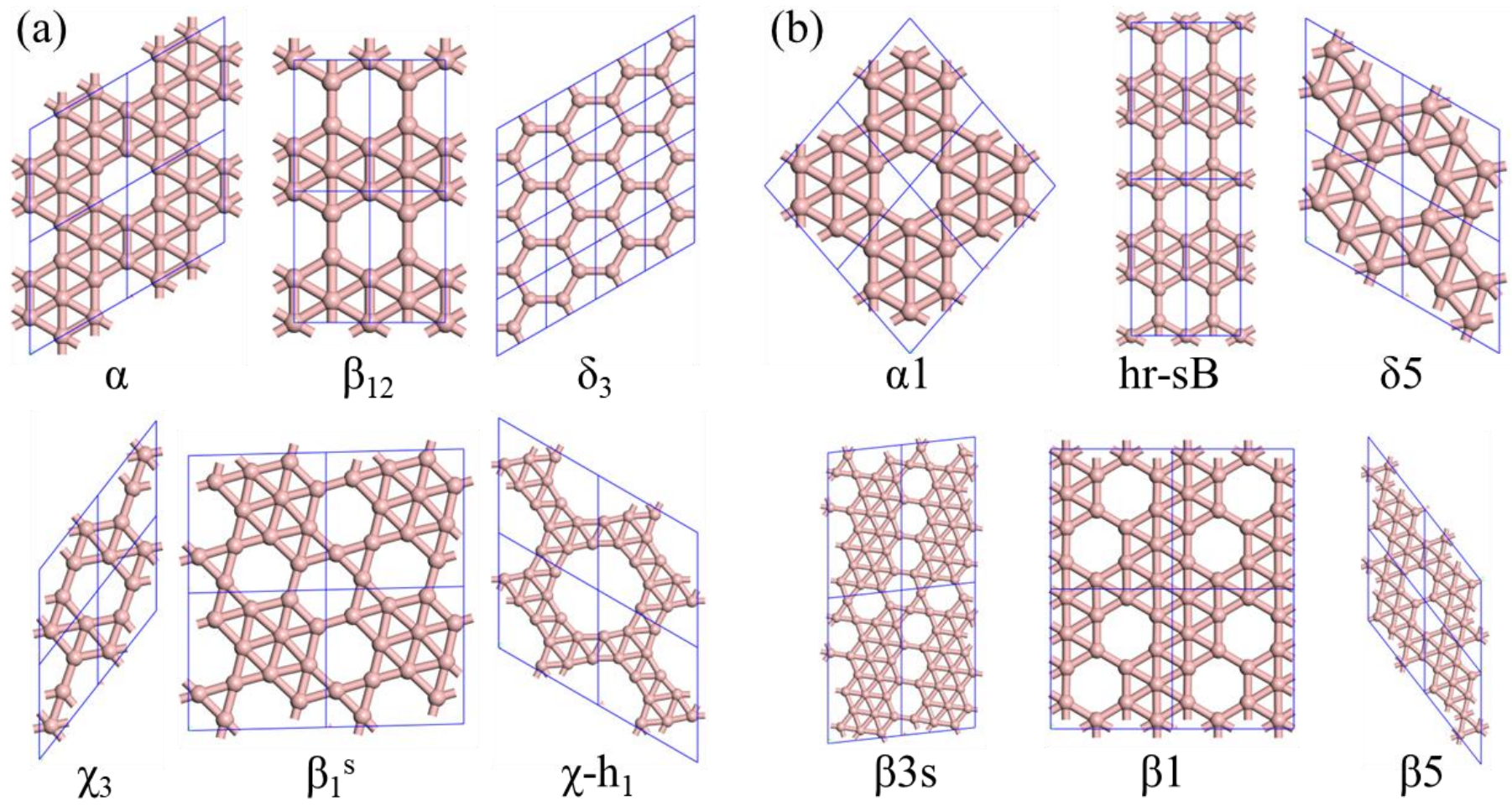

FIG S1. (a) The fully relaxed structures used for optimizing the TB parameters (b) The fully relaxed structures used for testing the transferability of the TB parameters.

## 2. Tight-Binding Parameters within $sp^3$ basis.

Table S1. Tight-Binding Parameters within $sp^3$ basis for Borophene. The $V$ and $E_{onsite}$ parameters are in eV and the $S$ parameters are dimensionless.

| | | ***ssσ*** | ***spσ*** | ***ppσ*** | ***ppπ*** |
|---|---|---|---|---|---|
| ***t*** | $V$ | -5.335 | 4.643 | 3.210 | -3.127 |
| | $q_1$ | 1.891 | 2.221 | 2.130 | 2.738 |
| ***s*** | $S$ | 0.119 | -0.097 | 0.005 | 0.123 |
| | $q_2$ | 1.463 | 1.433 | 0.001 | 2.410 |
| $\boldsymbol{E_{onsite}}$ | | $E_s$=-7.863 | | $E_p$=-2.609 | |

## 3. Band structures of borophenes for fitting and validating the TB parameters by VASP and the TB model with the $sp^3$ basis.

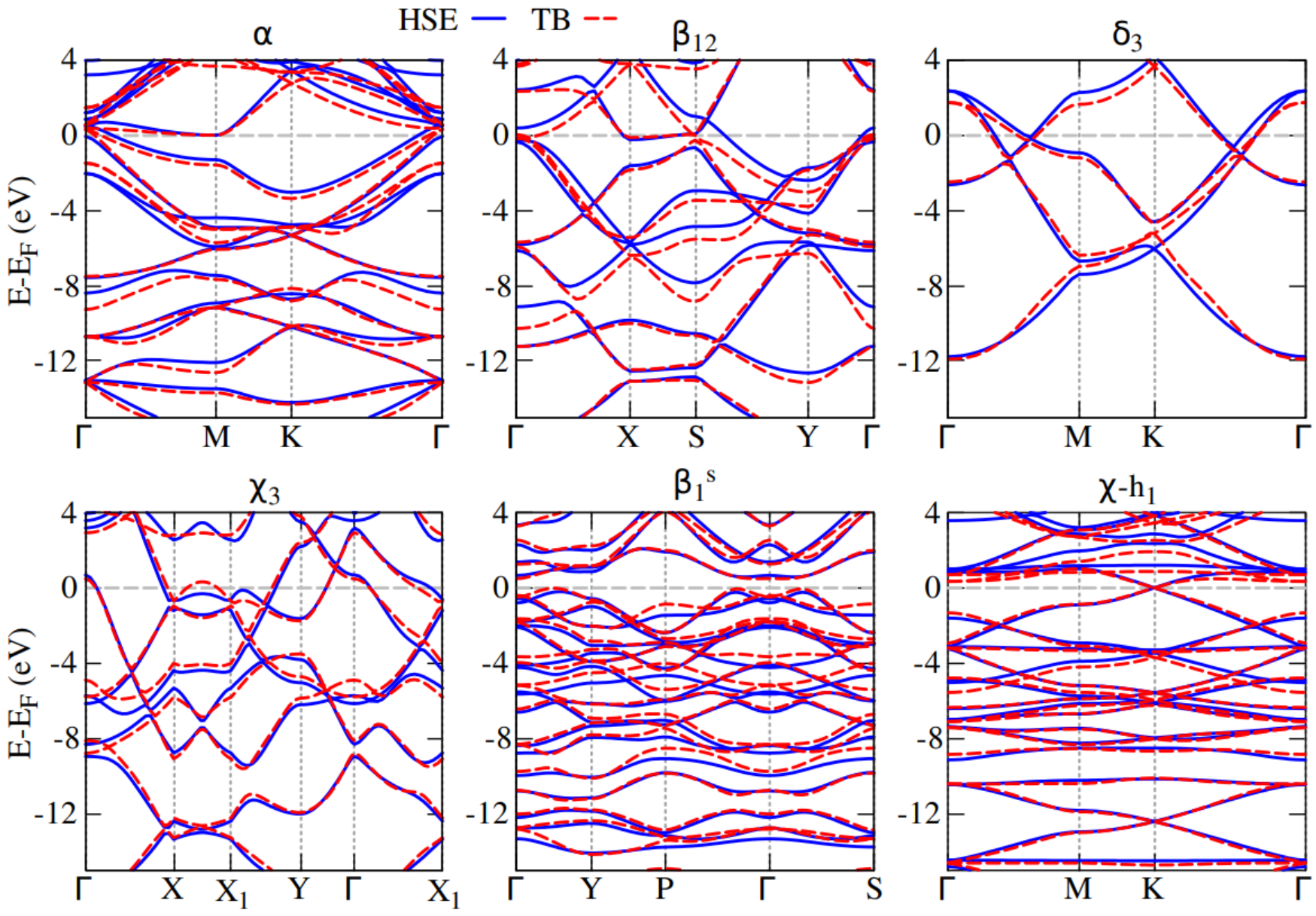

FIG S2. The calculated band structures of the six borophenes for fitting the TB parameters by VASP and the TB model with the $sp^3$ basis.

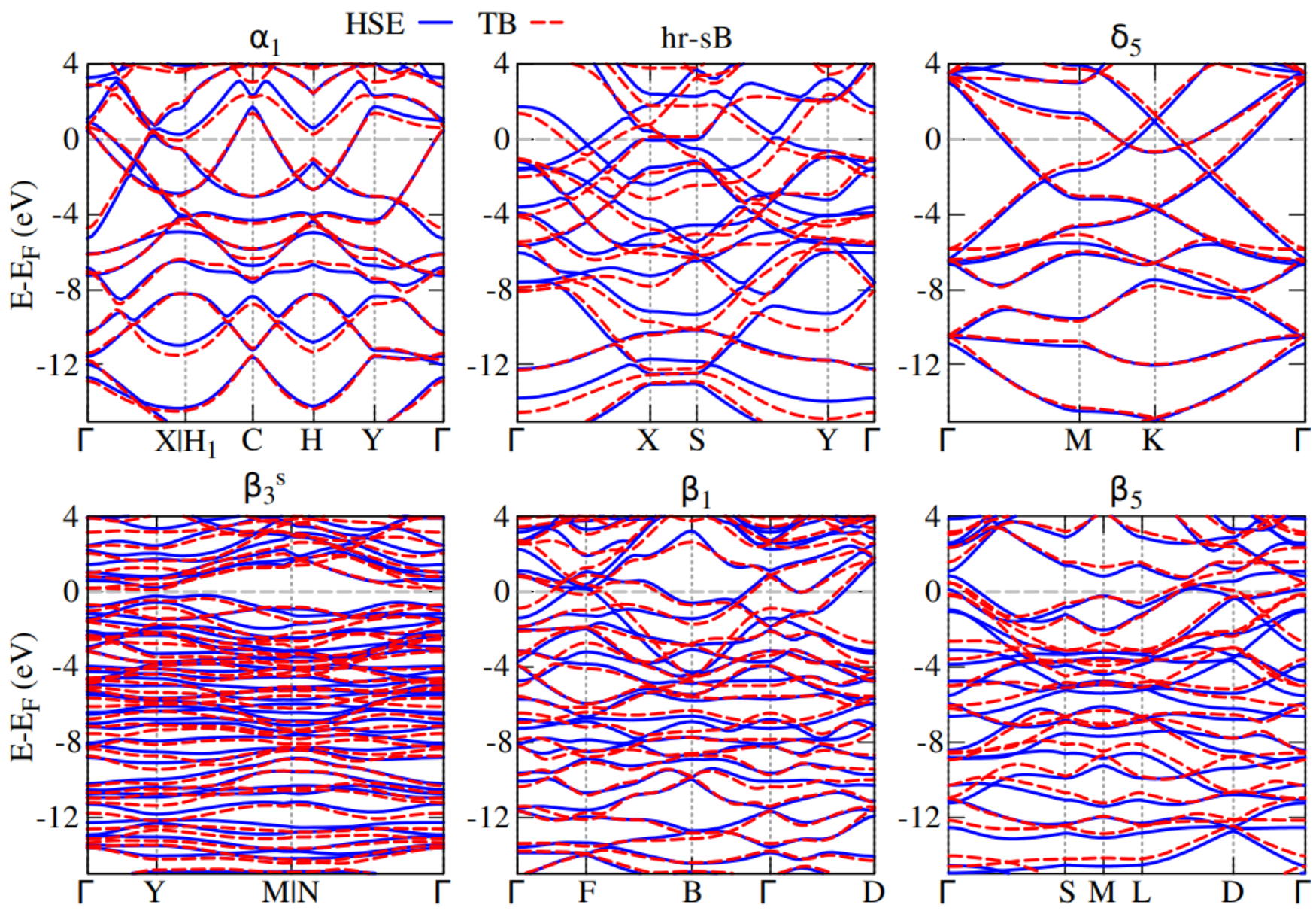

FIG S3. The calculated band structures of the six borophenes for testing the transferability of TB parameters by VASP and the TB model with the $sp^3$ basis.

4. Band structures of borophenes for fitting and validating the TB parameters by VASP and the TB model with the $sp^3d^5$ basis.

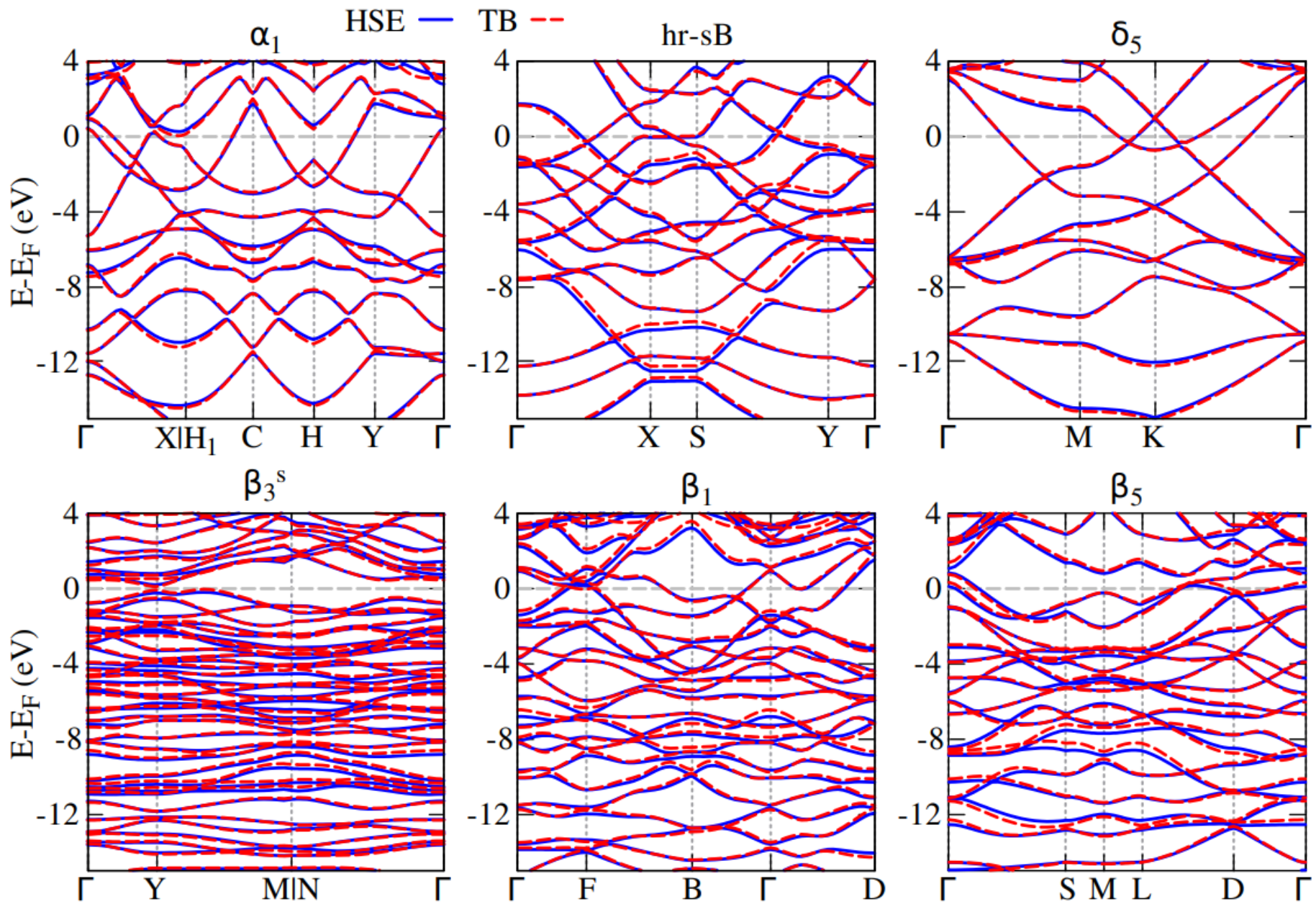

FIG S4. The calculated band structures of the six borophenes for testing the transferability of TB parameters by VASP and the TB model with $sp^3d^5$ basis.

## 5. Structure information of semimetallic borophenes

Table S2. Calculated lattice constants, atomic positions, and bond lengths of borophenes, density of vacancy η and Ratios of Number of Atoms with CN = 4, 5, 6 for borophenes.

| phase | *a* (Å) | *b* (Å) | γ (deg) | Atomic positions | bond lengths (Å) | *η* | ratios of CN (4:5:6) |
|---|---|---|---|---|---|---|---|
| 65-4-24-r68 | 8.756 | 8.464 | 90 | 8q (0.645,0.391,0.500)<br>8q (0.170,0.312,0.500)<br>4h (0.202,0.500,0.500)<br>4j (0.000,0.773,0.500) | 1.615–1.853 | 0.200 | 1:2:0 |
| 65-4-24-rx | 8.223 | 9.263 | 90 | 8q (0.595,0.680,0.500)<br>8q (0.708,0.164,0.500)<br>4h (0.602,0.000,0.500)<br>4j (0.500,0.158,0.500) | 1.570–1.750 | 0.200 | 2:3:1 |
| 47-5-14 | 5.776 | 8.412 | 90 | 4z (0.245,0.293,0.500)<br>4z (0.731,0.099,0.500)<br>2n (0.000,0.197,0.500)<br>2n (0.000,0.600,0.500)<br>2p (0.500,0.221,0.500) | 1.594–1.762 | 0.300 | 6:1:0 |
| 189-5-23 | 11.534 | 11.534 | 120 | 6k (0.255,0.496,0.500)<br>6k (0.430,0.595,0.500)<br>6k (0.084,0.659,0.500)<br>2d (0.333,0.667,0.500)<br>3g (0.496,0.000,0.500) | 1.568–1.754 | 0.521 | 15:6:2 |
| 174-6-16 | 8.856 | 8.856 | 120 | 3k (0.485,0.602,0.500)<br>3k (0.549,0.824,0.500)<br>3k (0.903,0.871,0.500)<br>3k (0.056,0.316,0.500)<br>3k (0.766,0.948,0.500)<br>1d (0.333,0.667,0.500) | 1.569–1.790 | 0.429 | 9:6:1 |
| 174-6-18 | 8.959 | 8.959 | 120 | 3k (0.964,0.695,0.500)<br>3k (0.033,0.912,0.500)<br>3k (0.137,0.333,0.500)<br>3k (0.224,0.546,0.500)<br>3k (0.827,0.759,0.500)<br>3k (0.116,0.653,0.500) | 1.587–1.865 | 0.357 | 3:2:1 |
| 51-10-36 | 13.277 | 8.983 | 90 | 4j (0.184,0.510,0.500)<br>4j (0.318,0.827,0.500)<br>4j (0.371,0.996,0.500)<br>4j (0.425,0.162,0.500)<br>4j (0.490,0.315,0.500)<br>4j (0.630,0.667,0.500)<br>4j (0.885,0.335,0.500)<br>4j (0.442,0.835,0.500)<br>2f (0.750,0.023,0.500)<br>2f (0.250,0.359,0.500) | 1.590–1.796 | 0.250 | 7:2:0 |
| 175-4-24 | 10.342 | 10.342 | 120 | 6k (0.290,0.291,0.500)<br>6k (0.369,0.469,0.500)<br>6k (0.630,0.099,0.500)<br>6k (0.819,0.638,0.500) | 1.600–1.778 | 0.351 | 1:0:0 |

6. Phonon spectrums and AIMD simulation results of semimetallic borophenes.

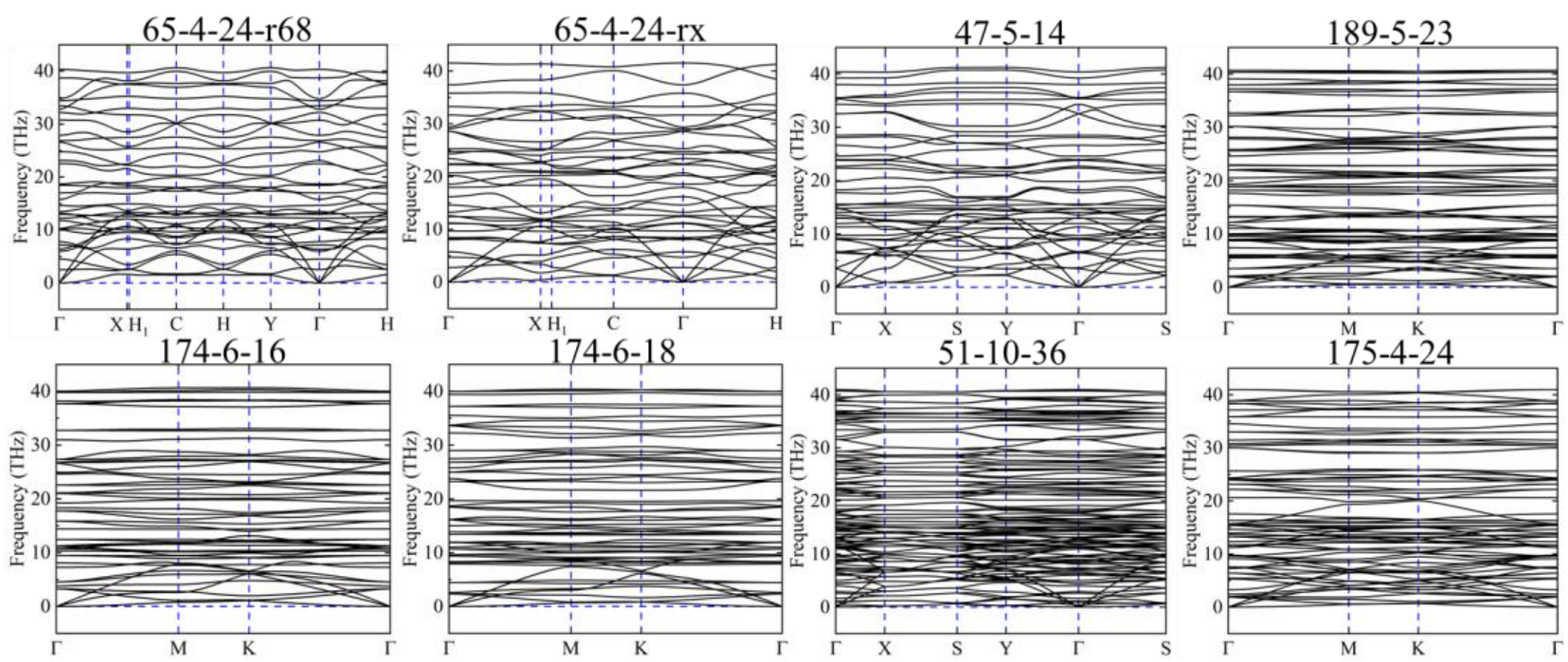

FIG S5. The calculated phonon band structure of 65-4-24-r68, 65-4-24-rx, 47-5-14, 189-5-23, 174-6-16, 174-6-18, 51-10-36 and 175-4-24.

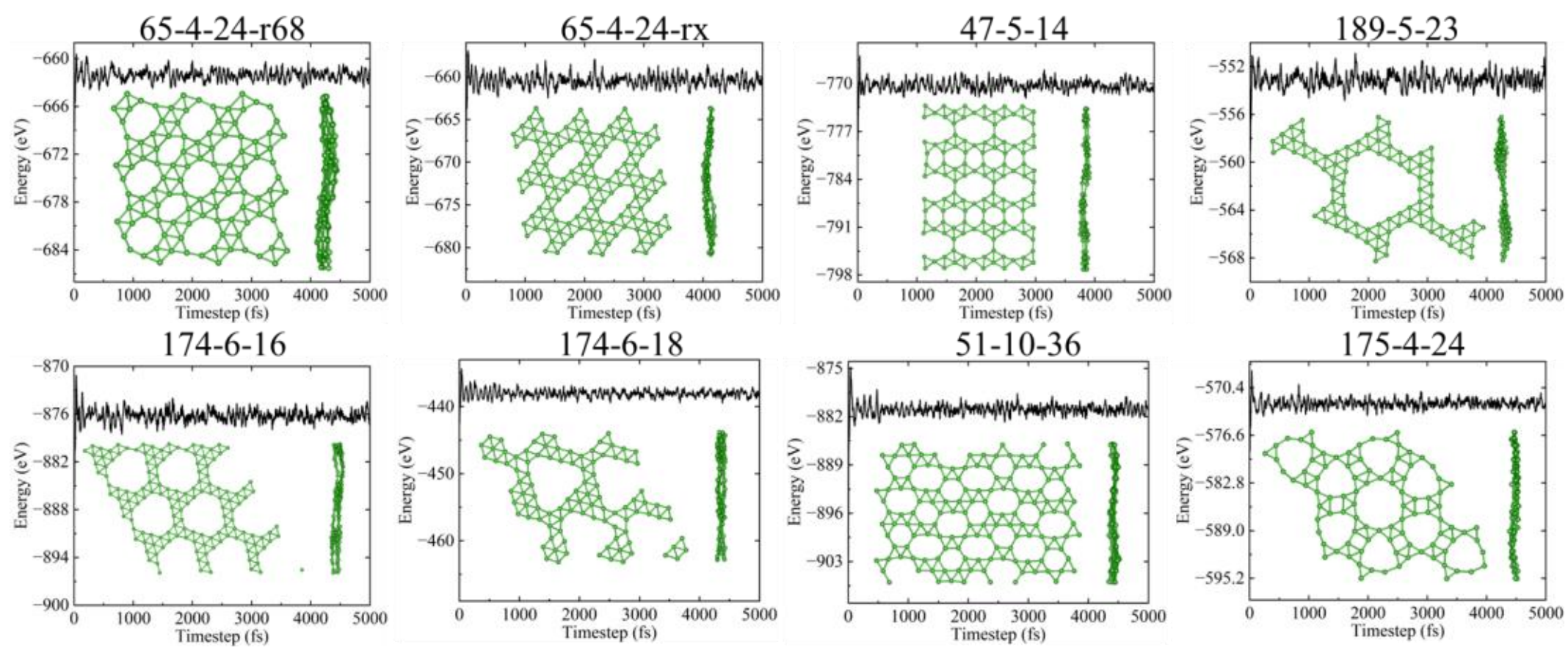

FIG S6. The total energy oscillations during the AIMD simulation at 500 K and the structural snapshots after 5 ps of AIMD simulation at 500K for 65-4-24-r68, 65-4-24-rx, 47-5-14, 189-5-23, 174-6-16, 174-6-18, 51-10-36 and 175-4-24.